\documentclass[twocolumn]{aastex701}

\usepackage{amsmath}

\accepted[Accepted for publication in ApJ]{}

\graphicspath{{./}}

\begin{document}

\title{Magnetic Flux Eruptions in Magnetically Arrested Disks: Magnetic Reconnection and Bubble Expansion}

\author[orcid=0000-0001-6545-4802]{Guoqiang Zhang}
\affiliation{Center for Astronomy and Astrophysics and Department of Physics, Fudan University, Shanghai 200438, China}
\email{guoqiangzhang@fudan.edu.cn}

\correspondingauthor{Feng Yuan}
\author{Feng Yuan}
\affiliation{Center for Astronomy and Astrophysics and Department of Physics, Fudan University, Shanghai 200438, China}
\email[show]{fyuan@fudan.edu.cn}

\begin{abstract}
  Using a high-resolution GRMHD simulation of a magnetically arrested disk, we investigate the physical mechanism driving magnetic-flux eruption.
  The horizon-threading magnetic flux declines mainly in off-equatorial regions, whereas localized hot spots associated with magnetic reconnection appear near the equatorial plane around the onset of flux decline.
  A non-axisymmetric, magnetically supported low-density structure, which we identify as a magnetic bubble, expands radially and azimuthally within a pre-existing low-density channel in the filamentary accretion flow.
  Based on these findings, we propose a physical picture linking reconnection, bubble expansion, and horizon magnetic flux decline.
  The reconnection occurs in the pre-existing low-density channel, and may reconfigure the field and promote bubble expansion.
  Magnetic-pressure gradients provide broadly distributed outward support for the expansion, while magnetic tension contributes a localized outward force near the equator.
  Bubble expansion may limit gas replenishment, helping to maintain the low-density region and reduce the total pressure near the equator.
  This pressure reduction may facilitate equatorward electromagnetic energy transport.
  In the simulation, the polar electromagnetic energy flux reverses from poleward to equatorward near the onset of the horizon-flux decline. The equatorward transport is concentrated in the bubble sector and carries electromagnetic energy from the off-equatorial magnetosphere to the equatorial region. This energy supply may sustain localized reconnection and plasma heating as the eruption develops.
\end{abstract}


\section{Introduction}
Magnetically arrested disks (MADs) are widely regarded as a  promising accretion mode of black hole hot accretion flows, which is associated with the launching of  relativistic jets~\citep{Narayan2003PASJ...55L..69N,Igumenshchev2008ApJ...677..317I,Yuan2014ARA&A..52..529Y}.
In this mode, large-scale poloidal magnetic flux is carried inward with the accretion flow and accumulated at the horizon, until the magnetic pressure near the black hole becomes strong enough to impede the inflow.
The accumulated flux supplies the ordered field required for efficient extraction of black hole rotational energy through the Blandford--Znajek mechanism~\citep{Blandford1977MNRAS.179..433B}.
It has been shown that the nearby supermassive black holes Sagittarius~A* and M87 may operate in this regime~\citep{EventHorizonTelescopeCollaboration2021ApJ...910L..13E,EventHorizonTelescopeCollaboration2022ApJ...930L..16E,Yuan2022ApJ...924..124Y}.

Much of the recent progress on MADs comes from general-relativistic magnetohydrodynamic (GRMHD) simulations~\citep{Tchekhovskoy2011MNRAS.418L..79T,Tchekhovskoy2012MNRAS.423L..55T,McKinney2012MNRAS.423.3083M,Ressler2017MNRAS.467.3604R,Chael2019MNRAS.486.2873C,Liska2020MNRAS.494.3656L,Yang2021ApJ...914..131Y,Begelman2022MNRAS.511.2040B,Narayan2022MNRAS.511.3795N,Zhang2024ApJ...962..135Z}, which self-consistently capture the nonlinear evolution of accretion, magnetic flux transport, and jet launching in the strong-gravity regime.
\citet{Tchekhovskoy2011MNRAS.418L..79T} performed a GRMHD simulation with spin $a = 0.99$ and found that the disk operates in the MAD state once the dimensionless magnetic flux reaches about 50.
They also found that the jet efficiency can exceed 100\%, which suggests that the jet extracts the rotational energy from the black hole.
\citet{Yang2021ApJ...914..131Y} used 3D GRMHD simulations to examine how black hole spin and magnetic-flux state affect wind and jet properties such as mass flux and velocity, whose properties are important for the study of AGN feedback \citep[e.g.,][]{2026SciA...12d6394H}.
\citet{Narayan2022MNRAS.511.3795N} performed several GRMHD simulations with different spins, and found that the dimensionless magnetic flux and the disk jet geometry depend strongly on the black hole spin.

Simulations further reveal that the accumulated flux in MAD systems is not steady; instead it is released in quasi-periodic bursts.
Such flux eruptions appear in a broad range of simulations~\citep{Porth2021MNRAS.502.2023P,Ripperda2022ApJ...924L..32R,Chatterjee2022ApJ...941...30C,Nalewajko2024A&A...692A..37N} and have been invoked to interpret the infrared and X-ray flare activity of Sgr~A*~\citep{GRAVITYCollaboration2018A&A...618L..10G,GRAVITYCollaboration2020A&A...638A...2G,Porth2021MNRAS.502.2023P,Scepi2022MNRAS.511.3536S,Cemeljic2022ApJ...933...55C,Grigorian2024MNRAS.530.1563G,Vos2024A&A...689A.112V,Antonopoulou2025A&A...696A..10A}.
\citet{Yang2024SciA...10N3544Y} proposed that a flux eruption event can strongly perturb the magnetic field lines of jet and trigger magnetic reconnection within the jet.
It has been shown in this work that radiation from electrons accelerated by this reconnection process successfully accounts for radiation of the jet, including the elongated structure, the limb-brightening morphology, and the jet width as a function of distance, supplying strong evidence for the first time that the jet produced by the BZ mechanism can really explain the complicated jet radiations.

Although flux eruptions have been widely identified in simulations of MAD disks, their physical mechanism remains unclear.
Existing studies support an important role for magnetic reconnection in MAD flux eruptions.
Simulations have identified magnetic reconnection in the equatorial region during a magnetic-flux eruption~\citep[e.g.,][]{Ripperda2022ApJ...924L..32R}.
The measured rate was consistent with the horizon-flux decay timescale, supporting a direct connection between reconnection and magnetic flux loss.
The role of reconnection in horizon-flux decay has also been examined in black hole magnetospheres~\citep{Bransgrove2021PhRvL.127e5101B} and discussed in a resolution study of MAD simulations~\citep{Salas2024MNRAS.533..254S}.
A complementary analysis to GRMHD simulation data of MAD revealed that current sheets are concentrated near the equatorial plane and along the jet boundary~\citep{Lin2024MNRAS.531.3136L}.
\citet{Nalewajko2024A&A...692A..37N} proposed that density gaps produced by chaotic magnetic disconnection activate a pre-existing current layer, thereby triggering reconnection and a flux eruption.
These studies establish magnetic reconnection as a central element of flux eruption.
However,
the relation between the horizon-flux decline, magnetic reconnection, and the formation and expansion of magnetic bubbles is unresolved.

In this paper, we investigate the evolution of a flux eruption using high-resolution GRMHD simulations of black hole accretion flows, with the aim of clarifying the underlying physical mechanisms. We focus on two characteristic features of the eruption---the horizon flux decline and the emergence of an expanding magnetic bubble---and examine how they are dynamically connected.
We propose a physical picture that connects the horizon-flux decline, magnetic-bubble expansion, and magnetic reconnection within a single eruption event.
The paper is organized as follows.
Section~\ref{sec:simulation} describes the simulation setup and diagnostics.
We present the phenomenology of the eruption in Section~\ref{sec:results}.
In Section~\ref{sec:mechanism}, we propose a physical picture of the flux eruption and then examine it through analyses of the dynamics and energy transport.
We summarize our conclusions in Section~\ref{sec:discussion}.

\section{Simulation Setup and Diagnostics} \label{sec:simulation}
\subsection{Simulation setup}
We analyze the simulation presented by \citet{Yang2024SciA...10N3544Y}; the full numerical setup is described there, and we summarize only the ingredients relevant to the present work.
The simulation is performed with Athena++~\citep{Stone2020ApJS..249....4S,White2016ApJS..225...22W} using static mesh refinement.
The simulation uses horizon-penetrating spherical Kerr--Schild coordinates.
We focus on the high-resolution simulation of a MAD around a black hole with $a = 0.98$, referred to as MAD98.
The effective final resolution is $N_r \times N_\theta \times N_\phi = 1408\times 512\times 256$.
The torus is initialized according to \citet{Fishbone1976ApJ...207..962F} with an inner radius of $r_{\mathrm{in}} = 40.5 r_g$ and a radius of maximum pressure at $r_{\mathrm{peak}} = 80 r_g$.
The adiabatic index is set to $\Gamma = 4/3$, and the ratio of gas pressure to magnetic pressure $\beta = p_{\mathrm{gas}} / p_{\mathrm{B}}$ is normalized to $\beta_{\mathrm{min}} = 0.1$.
The simulation runs until $t = 40000 t_g$, by which time the inflow-outflow equilibrium radius has extended to $r = 80 r_g$.

\subsection{Basic Quantities and Averaging Conventions}
The stress--energy tensor of an ideal magnetohydrodynamic fluid is
\begin{equation}
  \label{eq:stress-energy}
  T^\mu_{\ \nu} = (\rho + u + p_{\mathrm{gas}} + b^2) u^\mu u_\nu - b^\mu b_\nu + p_{\rm tot} \, g^\mu_{\ \nu},
\end{equation}
where $u^\mu$ is the fluid four-velocity, $\rho$ is the rest mass density, $u$ is the internal energy density, $p_{\rm gas}$ is the gas pressure, $b^\mu$ is the magnetic four-vector with $b^2 \equiv b^\mu b_\mu$, and $p_{\rm tot} = p_{\rm gas} + p_{\rm mag}$ is the total pressure, with $p_{\rm mag}=b^2/2$.
We also use $T^\mu_{\mathrm{EM}\,\nu} = b^2 u^\mu u_\nu - b^\mu b_\nu + p_{\rm mag} \, g^\mu_{\ \nu}$ for the electromagnetic part of the stress--energy tensor.

To characterize the accretion state at the horizon, we measure the mass accretion rate $\dot{M}$, the magnetic flux $\Phi_B$, and the dimensionless magnetic flux $\phi_B$:
\begin{equation}
  \label{eq:mad-factor}
  \begin{aligned}
    \dot{M} & = - \int_\phi \int_\theta \sqrt{-g}\,\rho\,u^r\,d\theta\,d\phi      \\
    \Phi_B  & = \int_\phi \int_\theta \sqrt{-g}\,\lvert B^r\rvert\,d\theta\,d\phi \\
    \phi_B  & = \frac{\Phi_B}{\sqrt{\dot{M}}}.
  \end{aligned}
\end{equation}
Here, $u^r$ is the radial component of the four-velocity, $B^r$ is the radial component of the magnetic field, and $g$ is the determinant of the metric.
The integrals are performed over the spherical surface at $r = r_h$, where $r_h$ is the horizon radius.

We also consider the $\theta$-average, $\phi$-average, and their combined $\theta$-$\phi$-average of various quantities at the horizon, which are defined as
\begin{equation}
  \label{eq:average-formula}
  \begin{aligned}
    \langle X \rangle_{\theta}      & = \frac{\int_{\theta_-}^{\theta_+} \sqrt{-g}\,X\,d\theta}
    {\int_{\theta_-}^{\theta_+} \sqrt{-g}\,d\theta}                                                                  \\
    \langle X \rangle_{\phi}        & = \frac{\int_0^{2\pi} \sqrt{-g}\,X\,d\phi}
    {\int_0^{2\pi} \sqrt{-g}\,d\phi}                                                                                 \\
    \langle X \rangle_{\theta,\phi} & = \frac{\int_{\theta_-}^{\theta_+} \int_0^{2\pi} \sqrt{-g}\,X\,d\phi\,d\theta}
    {\int_{\theta_-}^{\theta_+} \int_0^{2\pi} \sqrt{-g}\,d\phi\,d\theta}.
  \end{aligned}
\end{equation}
Here, $X$ is an arbitrary quantity, and $\theta_\pm$ are the limits of the $\theta$-average.
The $\theta$-average is performed over a narrow band around the equator to suppress short-wavelength variations in the disk region, while the $\phi$-average is performed over the full azimuthal range.

\section{Phenomenology of the Flux Eruption} \label{sec:results}
\subsection{Onset and Global Evolution} \label{subsec:general}
\begin{figure}
  \centering
  \includegraphics[width=0.95\linewidth]{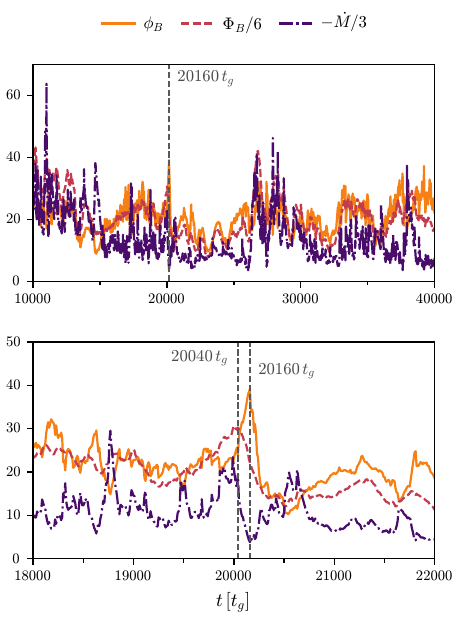}
  \caption{Time evolution of the mass accretion rate $\dot{M}$, magnetic flux $\Phi_B$, and normalized magnetic flux $\phi_B$ at the horizon.
    The three quantities are scaled for visual comparison.
    The top panel shows the long-term evolution from $t=10000$ to $40000\,t_g$.
    The bottom panel zooms into the representative eruption studied in this paper.
    In the bottom panel, the vertical dashed lines mark the onset of the sharp flux decrease at $t=20040\,t_g$ and the later peak of the dimensionless magnetic flux at $t=20160\,t_g$.}
  \label{fig:mdot-phib}
\end{figure}

The MAD state is characterized by large fluctuations in the magnetic flux threading the horizon.
We examine the time evolution of $\dot{M}$, $\Phi_B$, and $\phi_B$ at the horizon to locate the onset of the flux eruption.
Figure~\ref{fig:mdot-phib} shows the results.
We focus on a representative eruption event in which the dimensionless magnetic flux reaches a local maximum at $t = 20160 \, t_g$.
A closer view of the interval $18000$--$22000\,t_g$ is shown in the bottom panel.
The sharp decline in $\Phi_B$ begins at $t=20040\,t_g$, while the mass accretion rate $\dot{M}$ peaks at $t \sim 20000 \, t_g$, nearly simultaneously with the maximum of $\Phi_B$.
Thereafter, $\dot{M}$ falls more rapidly than $\Phi_B$.
Because $\phi_B = \Phi_B/\sqrt{\dot{M}}$, the dimensionless flux continues to rise during this interval even though the absolute magnetic flux is already declining.
Based on physical considerations, we identify $t = 20040 \, t_g$ as the onset of the flux eruption and use it  as a reference time for analyzing the subsequent evolution of the event.

\subsection{Latitude Dependence of the Flux Drop} \label{subsec:madfactordistribution}
\begin{figure*}
  \centering
  \includegraphics[width=1.0\linewidth]{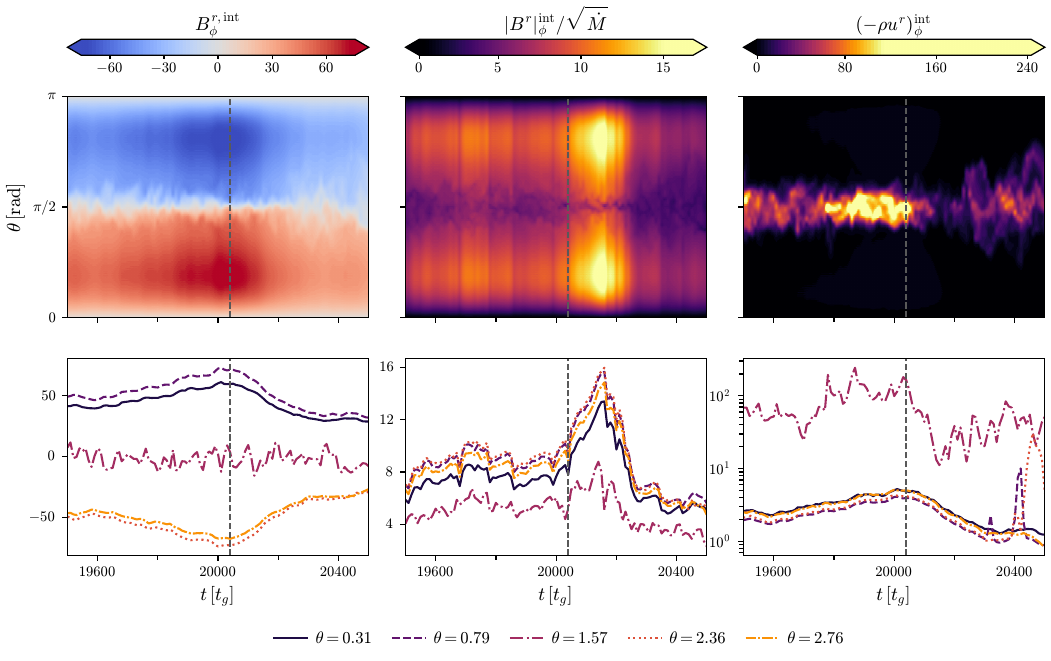}
  \caption{Time evolution of the $\phi$-integrated radial magnetic field $B^{r,\mathrm{int}}_\phi$, normalized magnetic flux $\lvert B \rvert^{r,\mathrm{int}}_\phi / \sqrt{\dot{M}}$, and mass-inflow flux $\left(-\rho u^r\right)^{\mathrm{int}}_\phi$ at the horizon.
  The top panels show the $\theta$--time distributions, while the bottom panels show the corresponding evolution at selected $\theta$ values.
  The vertical dashed line marks the onset of the magnetic-flux decline at $t = 20040\,t_g$.
  The radial magnetic field and normalized magnetic flux are concentrated mainly in regions away from the equatorial plane in both hemispheres and decrease after the eruption onset, whereas the mass-inflow flux is concentrated near the equatorial disk.
  These distributions show that the sharp magnetic-flux drop occurs mainly in off-equatorial regions rather than near the equator.}
  \label{fig:horizon-evolution}
\end{figure*}

To localize the eruption on the horizon, we examine $\phi$-integrated quantities at the horizon as functions of polar angle $\theta$ and time, derived from Equation~\ref{eq:average-formula} but without the denominator.
Figure~\ref{fig:horizon-evolution} shows the $\phi$-integrated radial magnetic field $B^{r,\mathrm{int}}_\phi$, normalized magnetic flux $\lvert B \rvert^{r,\mathrm{int}}_\phi / \sqrt{\dot{M}}$, and mass-inflow flux $\left(-\rho u^r\right)^{\mathrm{int}}_\phi$.
Here the superscript ``$\mathrm{int}$'' indicates that the quantity is integrated rather than averaged.
The top panels show the evolution across polar angle $\theta$, while the bottom panels show the evolution at selected $\theta$ values for reference.

The $\phi$-integrated radial field $B^{r,\mathrm{int}}_\phi$ has opposite signs across the equatorial plane, as expected from the large-scale field geometry.
Its magnitude is concentrated in two off-equatorial bands, $\theta \sim 0.5$--$1.0$ and $\theta \sim 2.2$--$2.8$.
In both bands, the radial field strengthens from $t \approx 19800\,t_g$ to $t = 20040\,t_g$ and then declines rapidly.
The selected-$\theta$ curves in the lower left panel illustrate this more clearly: the equatorial field remains weak and nearly unchanged, whereas the field in these off-equatorial regions is always much stronger.
It grows during this same interval and then falls rapidly.
The normalized flux, $\lvert B \rvert^{r,\mathrm{int}}_\phi / \sqrt{\dot{M}}$, follows the same latitude structure: it is strongest in these off-equatorial regions and remains weak near the equator throughout the evolution.

The $\phi$-integrated mass-inflow flux $\left(-\rho u^r\right)^{\mathrm{int}}_\phi$ shows a markedly different distribution from the magnetic field.
It is concentrated near the equator, as expected if mass is accreted primarily through the disk.
There is a clear brightening before the onset of the flux eruption, indicating that a substantial amount of material is being accreted.

\begin{figure*}
  \centering
  \includegraphics[width=\textwidth]{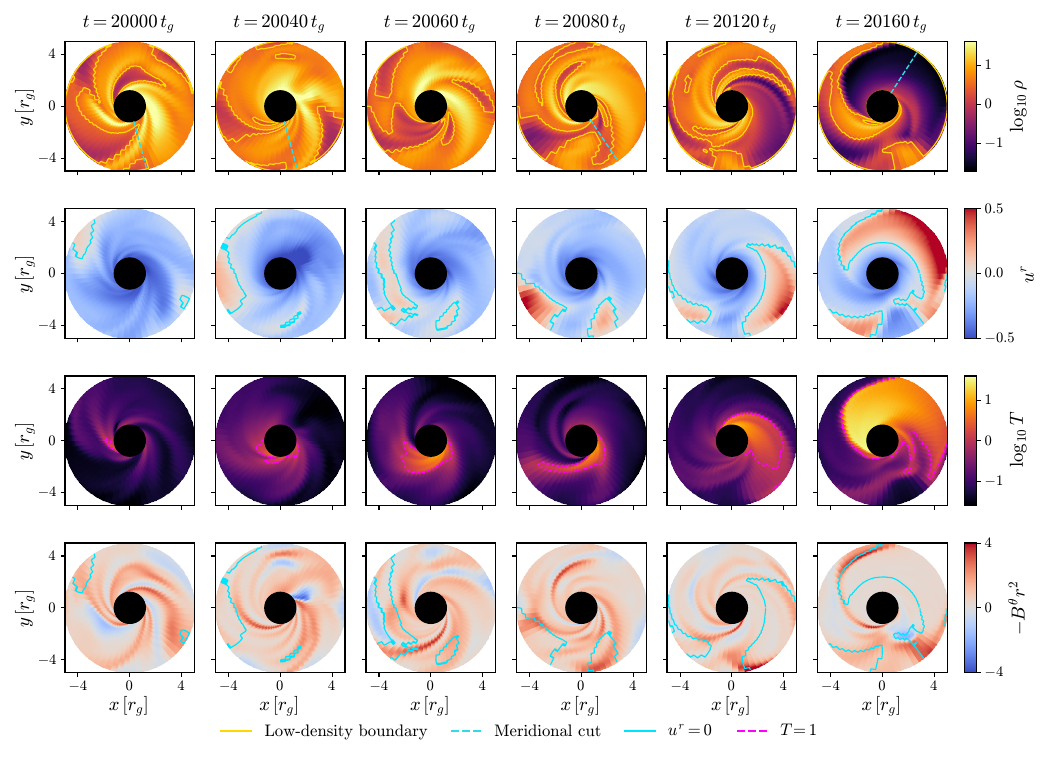}
  \caption{Equatorial slices of density $\rho$ (top), radial four-velocity $u^r$ (second row), and temperature $T=p_{\rm gas}/\rho$ (third row), and the radially weighted polar magnetic-field component $-B^\theta r^2$ (bottom row) at representative stages of the flux eruption.
  All quantities are $\theta$-averaged over $\theta \in (\pi/2-0.05,\,\pi/2+0.05)$ using Equation~\ref{eq:average-formula}.
  Density and temperature are shown on logarithmic scales; all panels in each row share the same color scale.
  Thin gold curves outline the low-density regions discussed in Section~\ref{subsec:bubble-identification}.
  Cyan dashed radii mark the selected azimuths for the right-hand halves of the meridional cuts in Figure~\ref{fig:magnetic-bubble-meridional}.
  Solid cyan curves mark $u^r=0$ in the velocity and signed polar-field rows; the dashed magenta $T=1$ contour appears only in the temperature row.
   The low-density magnetic bubble is identifiable near $(x,y)\approx(2,-3)$ at $t=20080\,t_g$, co-rotates with the disk, and expands substantially by $t=20160\,t_g$.
  The bubble is associated with outward motion.
  A broad hot region is visible by $t=20120\,t_g$ and partially overlaps the bubble at $t=20160\,t_g$.}
  \label{fig:magnetic_bubble}
\end{figure*}

\begin{figure*}[t]
  \centering
  \includegraphics[width=\textwidth]{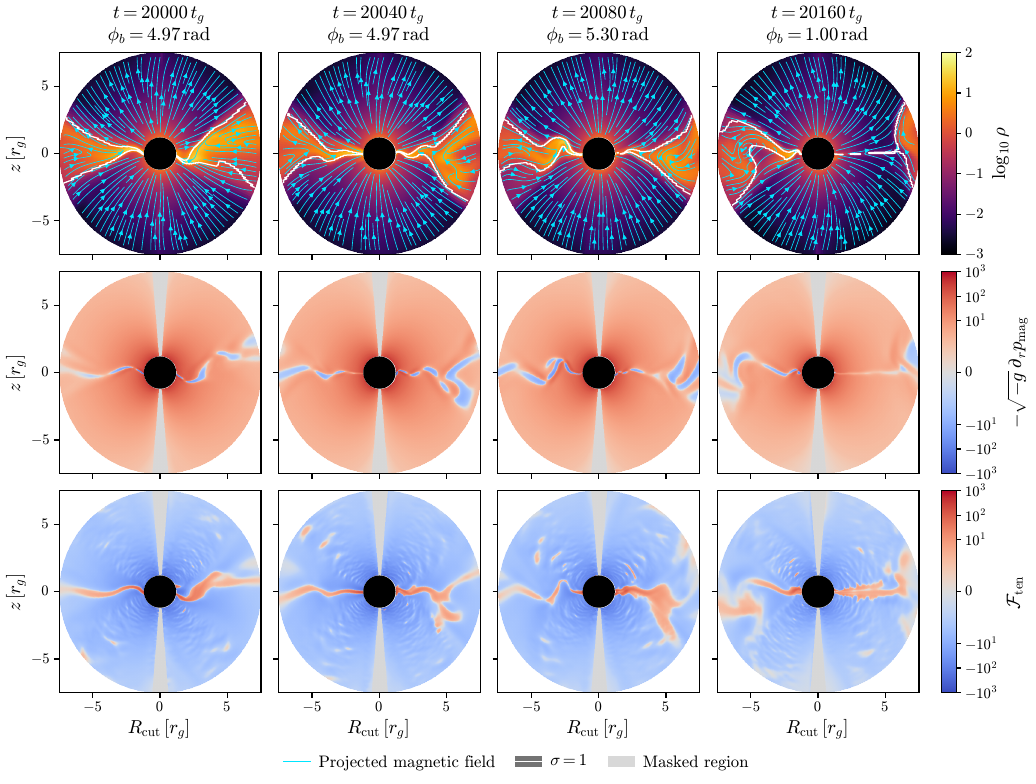}
  \caption{Meridional density and radial magnetic-stress cuts at $t=20000$, $20040$, $20080$, and $20160\,t_g$ (left to right).
  The rows show $\log_{10}\rho$ (top), the radial magnetic-pressure-gradient term $-\sqrt{-g}\,\partial_r p_{\rm mag}$ (middle), and the radial magnetic tension $\mathcal{F}_{\rm ten}$ defined in Equation~\ref{eq:conserved-equation} (bottom).
  In the top row, cyan streamlines show the direction of the magnetic field projected onto the meridional plane, and white contours mark $\sigma=1$.
  In each panel, the right half uses the $\phi$ direction indicated by the cyan dashed radius in Figure~\ref{fig:magnetic_bubble}, while the left half shows the antipodal $\phi$ direction.
  The first two columns use the same azimuth, while the last two show cuts through the magnetic bubble identified in Figure~\ref{fig:magnetic_bubble}, at an early stage ($t=20080\,t_g$) and after substantial expansion ($t=20160\,t_g$), respectively.
  The lower two rows share a symmetric logarithmic color scale.
  Positive values indicate radially outward contributions and negative values inward contributions on both sides.
      Gray polar strips mark regions with insufficient neighboring cells for the spatial filtering and derivatives.
      At $t=20160\,t_g$, the magnetic bubble extends  to approximately $6r_g$; a nearly vertical projected magnetic field appears close to its radial outer edge.}
  \label{fig:magnetic-bubble-meridional}
\end{figure*}

The evolution shown in Figure~\ref{fig:horizon-evolution} indicates that the sharp drop in magnetic flux mainly occurs in regions away from the equatorial plane in both hemispheres rather than near the equator.
However, previous studies suggest that magnetic reconnection during flux eruptions primarily occurs in the equatorial plane~\citep{Ripperda2022ApJ...924L..32R,Lin2024MNRAS.531.3136L,Nalewajko2024A&A...692A..37N,Jacquemin-Ide2025arXiv251025842J}.
Thus, the regions of magnetic-flux loss do not spatially coincide with those of magnetic reconnection, and a mechanism is needed to connect these two processes.

\subsection{Magnetic Bubble Emergence and Growth} \label{subsec:bubble-identification}

Another important signature of magnetic-flux eruptions is the emergence and expansion of magnetic bubbles near the event horizon.
Here, ``magnetic bubble'' refers to a non-axisymmetric, low-density, highly magnetized region intersecting the equatorial plane.
Magnetic pressure provides the main support within this region.
We examine its morphology and growth in this simulation.
Figure~\ref{fig:magnetic_bubble} shows equatorial slices of density $\rho$, radial four-velocity $u^r$, temperature proxy $T = p_{\rm gas} / \rho$, and the radially weighted polar magnetic-field component $-B^\theta r^2$.
To suppress short-wavelength variations around the midplane, we compute a $\theta$-average over a narrow band around the equator using Equation~\ref{eq:average-formula}, with the average taken over $\theta \in (\pi/2 - 0.05, \, \pi/2 + 0.05)$.
For brevity, we omit the $\langle \rangle_\theta$ label in this subsection and Figure~\ref{fig:magnetic_bubble}.

We first examine the equatorial density maps in the top row of Figure~\ref{fig:magnetic_bubble}.
Thin gold curves outline connected low-density regions defined by both $\rho<3$ and $\rho<0.3\max_\phi\rho(r,\phi)$; these numerical thresholds are chosen only to illustrate the morphology of the bubble, not to define a robust physical boundary.
Several low-density regions appear in the equatorial plane; we single out the low-density region near $(x,y)\approx(2,-3)$ at $t = 20080 \, t_g$.
 This structure is clearly identifiable at $t=20080\,t_g$, shortly after the onset of the flux eruption ($t = 20040\,t_g$).
By $t=20160\,t_g$, the low-density region has expanded in both the radial and azimuthal directions, becoming the dominant low-density feature in the near-horizon region.
We identify this expanding, magnetically supported low-density structure as the magnetic bubble associated with the flux eruption.
It  expands as the horizon flux decreases.
It later partially overlaps with the high-temperature region, further supporting this identification.

The radial-velocity maps give the kinematic counterpart.
Compared with the density maps, the low-density region largely overlaps the region with $u^r>0$ at $t=20080, 20120, 20160\,t_g$, consistent with the bubble expansion.
A weak outward-motion signature is already visible in the sector near $(x,y)\approx(0,-3)$ at $t=20040\,t_g$, close to the onset of the horizon-flux decline.
The density map shows a corresponding region of relatively low density, although it is not enclosed by the gold contours.

The temperature map (third row) shows that part of the magnetic bubble overlaps with a broad high-temperature region at $t=20160\,t_g$, consistent with the results of \citet{Nalewajko2024A&A...692A..37N}.
The dashed magenta $T=1$ contours show small heated patches near the black hole already at $20040\,t_g$, followed by a narrow heated arc at $t=20060$--$20080\,t_g$.
A much broader and hotter region is already visible at $t=20120\,t_g$ and occupies a larger area at $t=20160\,t_g$.
The early near-horizon hot patches lie in regions of relatively low density, consistent with localized reconnection in the low-density environment at the onset of the flux decline.

To examine the vertical magnetic field within and around the bubble, we show the radially weighted component normal to the equatorial plane, $-B^\theta r^2$, in the bottom row of Figure~\ref{fig:magnetic_bubble}.
The $r^2$ weighting highlights the polar-field distribution at larger radii.
By $t=20120$--$20160\,t_g$, enhanced signal appears at larger radii and follows part of the bubble's radial outer boundary, whereas much of its interior has a weaker polar component.
This contrast suggests that the vertical field is locally concentrated near the bubble's radial outer edge rather than in its interior.

To examine the magnetic-field geometry within and around the bubble, we show the density and magnetic-field directions in meridional slices in the top row of Figure~\ref{fig:magnetic-bubble-meridional}.
The selected azimuths are indicated by the cyan dashed radii in the density row of Figure~\ref{fig:magnetic_bubble}.
The right part in each panel shows the cut along the selected azimuth, while the left part shows the antipodal cut.
The $\sigma=1$ contours (white lines) provide a reference for the magnetization structure.
Here, $\sigma=b^2/\rho$ denotes the magnetization.
The cuts at $t=20000$ and $20040\,t_g$ sample the same azimuth, providing a comparison before and at the onset of the flux decline.
At $t=20040\,t_g$, the cut samples the sector containing the weak outward-motion signature near $(x,y)\approx(0,-3)$ in Figure~\ref{fig:magnetic_bubble}.
On the right side of this early cut, the projected field bends toward a nearly vertical orientation at larger radii.
At $t=20080$ and $20160\,t_g$, the selected cuts show a similar bending of the projected field toward a nearly vertical orientation near the bubble's radial outer edge, while the interior field is predominantly radial.
The near-vertical field at the bubble's outer edge is consistent with the geometry of ejected magnetic flux tubes reported in previous simulations \citep{Ripperda2022ApJ...924L..32R,Nalewajko2024A&A...692A..37N}.
In particular, \citet{Ripperda2022ApJ...924L..32R} showed that reconnection converts the field along the equatorial current sheet into vertical magnetic flux expelled from the reconnection layer.

\subsection{Key Constraints on the Eruption Mechanism}\label{subsec:key-constraints}

These results identify the spatial and temporal behavior to be connected in the eruption picture developed below.
Spatially, the sharp drop in horizon-threading magnetic flux occurs mainly in regions away from the equatorial plane in both hemispheres, whereas localized heating consistent with magnetic reconnection is concentrated  near the equatorial plane.
Temporally, the initial outward motion and localized hot patches are present near the onset of the flux decline; the bubble and heated region then grow as the eruption proceeds.
The following analysis examines the magnetic forces  contributing to the radial expansion and the energy transport linking off-equatorial field weakening to magnetic reconnection near the equatorial plane.

\section[Physical Picture of the Flux Eruption]{Physical Picture of the Flux Eruption} \label{sec:mechanism}

\subsection[Overview of the Eruption Scenario]{Overview of the Eruption Scenario}\label{subsec:mechanism-overview}

\begin{figure*}
  \centering
  \includegraphics[width=\textwidth]{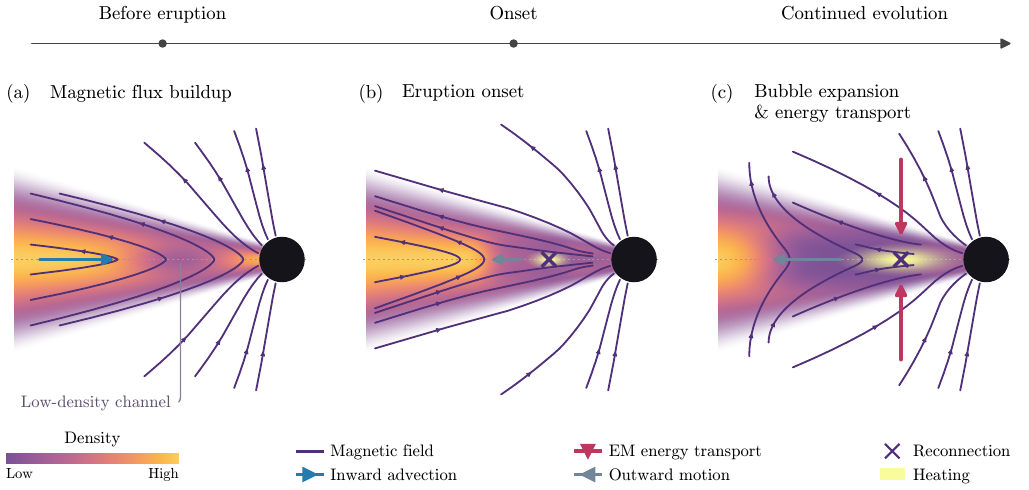}
  \caption{Proposed physical picture of the flux eruption, connecting localized magnetic reconnection, bubble expansion, and flux decline at the horizon.
    Each panel shows a local meridional section.
    (a) During the pre-eruption phase, magnetic flux is advected inward and accumulates at the horizon.
    A low-density channel is already present within the equatorial accretion flow.
    (b) Localized magnetic reconnection near the horizon  may reconfigure the field within the pre-existing low-density channel and  heat the plasma.
    Magnetic pressures and magnetic tension may drive the outward expansion of the bubble.
    (c) As the bubble expands, electromagnetic energy is transported from both hemispheres toward the equatorial reconnection region, as indicated by the red arrows.
    We further propose that outward motion may reduce gas replenishment and help preserve the low-density environment.
    The projected field is predominantly radial within the developed bubble and bends toward a near-vertical direction at its radial outer edge.
    The equatorward transport shown in panel (c) begins near the onset depicted in panel (b) and continues as the bubble expands.
    The color shading indicates density from low (purple) to high (orange-yellow).
    Purple curves show magnetic field lines; blue, gray, and red arrows show inward advection, outward motion, and equatorward electromagnetic-energy transport, respectively.
    The black circle represents the black hole.
    Magnetic field lines are depicted only outside the horizon.
    The purple cross marks the X-point, and the yellow region shows localized heating.}
  \label{fig:eruption-overview}
\end{figure*}

The constraints identified in Section~\ref{sec:results} motivate a physical picture that links magnetic reconnection near the equatorial plane, magnetic bubble expansion, and the decline of horizon-threading magnetic flux in off-equatorial regions.
We therefore propose the eruption scenario illustrated in Figure~\ref{fig:eruption-overview}, as a working model to be examined by the diagnostics presented below.
In this scenario, the horizon-threading magnetic flux builds up, while the filamentary equatorial flow contains low-density regions, consistent with interchange instability.
During this accumulation phase, the polar electromagnetic energy flux is directed away from the equatorial plane, transporting magnetic energy toward the off-equatorial regions.
This magnetic-flux buildup increases the magnetic pressure on both sides of the equatorial plane. The increased pressure may bring oppositely directed field lines closer together within the pre-existing low-density channels, facilitating localized reconnection.
Reconnection  can produce strongly curved magnetic field lines whose tension can pull the reconnected field outward.
Together with the magnetic-pressure gradients, this outward tension  may contribute to the expansion of the low-density channel into a magnetic bubble.
As reconnection proceeds and the bubble expands, the total pressure near the equatorial plane may decrease, facilitating equatorward electromagnetic energy transport from the surrounding off-equatorial regions.
This energy supply can sustain further reconnection and plasma heating during the decline of horizon-threading magnetic flux.

In the following three subsections, we examine this scenario through three complementary diagnostics: the pre-existing low-density channels in the near-horizon flow, the magnetic forces  contributing to the radial expansion, and the electromagnetic-energy transport connecting the weakening off-equatorial magnetic field to the equatorial bubble region.

\subsection[Filamentary Accretion and the Low-density Environment]{Filamentary Accretion and the Low-density Environment} \label{subsec:trigger}
\noindent

We first examine the near-horizon accretion flow to characterize the pre-existing low-density environment depicted in panel (a) of Figure~\ref{fig:eruption-overview}.
As shown in Figure~\ref{fig:magnetic_bubble},  the bubble develops within a filamentary equatorial flow containing pre-existing low-density regions.
This morphology is consistent with the non-axisymmetric accretion pattern expected in the MAD state, commonly associated with interchange instability~\citep{Igumenshchev2008ApJ...677..317I,Begelman2022MNRAS.511.2040B}.

To quantify this structure, we decompose the azimuthal density distribution in the near-horizon, near-equatorial flow into Fourier modes:
\begin{equation}
  \label{eq:fourier-mode}
  A_m(\theta) = \frac{1}{2\pi}\int_0^{2\pi} \rho(\theta,\phi)\, e^{-im\phi}\,d\phi,
\end{equation}
We average the mode amplitudes over $r_{\rm H}<r<5r_g$ and $\theta \in (\pi/2-0.05,\,\pi/2+0.05)$ with the weight $\sqrt{-g}\,dr\,d\theta$, and normalize them by the corresponding axisymmetric ($m=0$) component.

\begin{figure}[t]
  \centering
  \includegraphics[width=\linewidth]{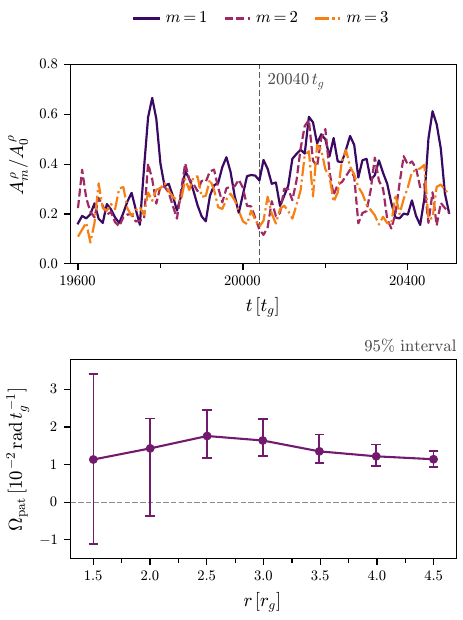}
  \caption{
    Top: Relative amplitudes $A_m^\rho/A_0^\rho$ of the $m=1,2,3$ density modes, averaged over $r_{\rm H}<r<5r_g$ and $|\theta-\pi/2|<0.05$.
    The dashed line marks $t=20040\,t_g$. Bottom: Characteristic density-pattern angular velocity at fixed radii.
    Error bars show 95\% block-resampling intervals.
    The Fourier amplitudes illustrate pre-existing non-axisymmetry; the correlation estimates quantify the pattern motion.}
  \label{fig:fourier-mode}
\end{figure}

Figure~\ref{fig:fourier-mode}(top panel) shows the resulting relative amplitudes $A_m^\rho/A_0^\rho$ for the low-order density modes.
The low-order modes maintain substantial amplitudes throughout the plotted interval.
These properties indicate that the filamentary structure is a persistent, evolving feature of the flow over the analyzed interval, already present before the eruption.
We also checked the Fourier modes of the magnetic-field strength $b^2$; the density and magnetic-field phases differ substantially, which is consistent with an interchange interpretation of the filamentary structure~\citep{McKinney2012MNRAS.423.3083M}.
The density patterns also exhibit prograde rotation, with characteristic pattern speeds of $\Omega_{\rm pat}\simeq0.0164$ and $0.0122\,\mathrm{rad}\,t_g^{-1}$ at $r\simeq3r_g$ and $4r_g$, respectively, as estimated from density correlations at fixed radii (Figure~\ref{fig:fourier-mode}, bottom panel).

The filamentary flow thus provides low-density channels before the eruption.
At the onset of the horizon-flux decline ($t=20040 t_g$), localized heating appears in relatively low-density regions near the black hole (Figure~\ref{fig:magnetic_bubble}).
The low-density channels are supported mainly by magnetic pressure, whereas gas pressure contributes more strongly in the dense filaments.
In our proposed picture, continued magnetic-flux accumulation strengthens the magnetic field above and below the equatorial plane.
This field can exert vertical magnetic compression on the equatorial flow, as discussed by \citet{Cao2011ApJ...737...94C}.
Within the low-density channels, the stronger magnetic field may allow oppositely directed field lines to come into closer contact under compression, facilitating localized reconnection.
The present diagnostics characterize the low-density environment but do not determine what triggers localized reconnection.
Magnetic compression is proposed here as a possible facilitating process, rather than an established trigger.

\subsection{Bubble Dynamics and Driving Forces} \label{subsec:dynamic-of-magnetic-bubble}

We next examine the radial force balance that supports the outward bubble growth shown schematically in panels (b) and (c) of Figure~\ref{fig:eruption-overview}.
To examine the magnetic forces driving bubble expansion, we begin with the $r$-component of the stress--energy conservation equation,
\begin{equation}
  \label{eq:conserved-equation}
  \begin{aligned}
    \partial_t (\sqrt{-g}\, T^t_r) ={} & - \partial_j (\sqrt{-g}\,(h + b^2)\, u^j u_r)                                \\
                                       & - \sqrt{-g}\, \partial_r p_{\rm tot}                                         \\
                                       & + \partial_j (\sqrt{-g}\, b^j b_r) - \sqrt{-g} b^\mu b_\nu \Gamma^\nu_{r\mu} \\
                                       & + \sqrt{-g}(h + b^2)\,u^\mu u_\nu \Gamma^\nu_{r\mu},
  \end{aligned}
\end{equation}
where $h = \rho + u + p_{\rm gas}$ is the fluid enthalpy, and $\Gamma^\nu_{r\mu}$ are the Christoffel symbols.
The terms on the right-hand side correspond to (from top to bottom): the divergence of the momentum flux carried by the fluid motion (enthalpy plus electromagnetic energy), the pressure-gradient term, the magnetic tension term ($\mathcal{F}_{\mathrm{ten}} = \partial_j (\sqrt{-g}\, b^j b_r) - \sqrt{-g} b^\mu b_\nu \Gamma^\nu_{r\mu}$), and the source term associated with spacetime curvature through the Christoffel symbols $\Gamma^\nu_{r\mu}$.
Here, the connection term associated with the anisotropic magnetic stress $-b^\mu b_\nu$ is grouped together with the magnetic tension contribution.

\begin{figure}[t]
  \centering
  \includegraphics[width=\linewidth]{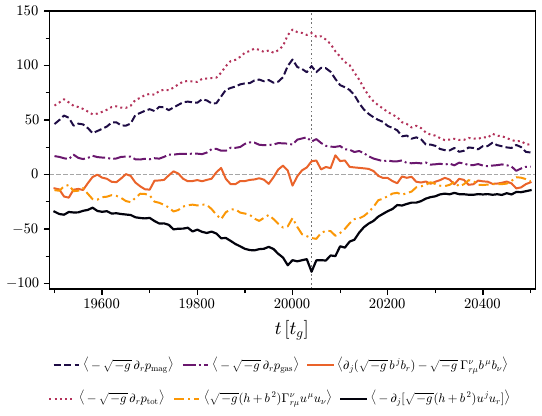}
  \caption{Evolution of the radial momentum terms in the radial stress--energy conservation equation (Eq.~\ref{eq:conserved-equation}), evaluated at $r\approx1.5\,r_g$ and averaged over $\theta \in (\pi/2 - 0.2, \pi/2 + 0.2)$ and $\phi \in (0, 2\pi)$ (Eq.~\ref{eq:average-formula}).
    The vertical line marks the onset of the flux eruption ($t = 20040\,t_g$).
    The total pressure gradient, which is dominated by the magnetic pressure $p_{\rm mag}=b^2/2$, increases during the accumulation phase and decreases rapidly after the eruption, while the gas pressure gradient varies less strongly.
    The magnetic tension becomes positive near the eruption onset.
    The negative momentum-flux divergence contributes substantially to the radial momentum balance.}
  \label{fig:r_dir_force_profile}
\end{figure}

Figure~\ref{fig:r_dir_force_profile} shows the radial momentum terms at $r\approx1.5\,r_g$, averaged over $\theta \in (\pi / 2 - 0.2, \pi / 2 + 0.2)$ and $\phi \in (0, 2\pi)$, which suppresses turbulent fluctuations (i.e., $\langle X \rangle_{\theta,\phi}$; see Equation~\ref{eq:average-formula}).
We omit the averaging subscripts in the text.
The vertical dashed line marks $t = 20040\,t_g$.
The total pressure gradient, dominated by the magnetic pressure, rises during the buildup phase and then falls after the eruption.
The gas pressure gradient is comparatively small.
The pressure gradients stay positive throughout, meaning that the pressure gradient consistently acts to push the accretion flow outward.
The angularly averaged magnetic tension term $\mathcal{F}_{\mathrm{ten}}$ is much smaller than the pressure-gradient terms.
Its value fluctuates before the eruption and becomes positive near the onset at $t=20040\,t_g$.
This indicates that the magnetic tension provides an additional outward radial force at the eruption onset.
Despite its small angular average, magnetic tension can be locally strong, as shown by the meridional force maps discussed below.

The advective geometric term $\langle \sqrt{-g} (h + b^2) \Gamma^\nu_{r\, \mu} u^\mu u_\nu \rangle$ is negative, corresponding to gravity in classical MHD.
This term also becomes less negative (i.e., rises toward zero) after the eruption, possibly because of the decrease in $b^2$.
The momentum-flux divergence is negative throughout the interval, with its largest magnitude near the eruption onset.
The negative momentum-flux divergence offsets the positive pressure-gradient term, leaving a small net contribution.

The lower two rows of Figure~\ref{fig:magnetic-bubble-meridional} resolve the spatial distribution of the radial magnetic-pressure-gradient and tension terms.
The magnetic-pressure-gradient term is outward over much of the meridional plane, with small inward contributions confined mainly to narrow bands near the equatorial flow, possibly associated with inhomogeneities in the magnetized flow.
Magnetic tension is predominantly inward away from the midplane, whereas its outward contribution is concentrated along the equatorial channel.
Near the equatorial plane, magnetic tension in parts of the bubble region acts radially outward and can be comparable to or larger than the magnetic-pressure-gradient term.
The angular average in Figure~\ref{fig:r_dir_force_profile} combines the bubble sector with the surrounding flow, diluting the strong outward tension concentrated in localized regions near the equatorial plane.
Together, Figures~\ref{fig:magnetic-bubble-meridional} and~\ref{fig:r_dir_force_profile}  show outward radial contributions from both magnetic-pressure gradients and localized magnetic tension, supporting their role in bubble expansion.

\subsection{Equatorward Electromagnetic Energy Transport} \label{subsec:equatorward-transport}\label{subsec:global-b-decline}

The preceding subsections examined the low-density environment and the magnetic forces  contributing to the radial expansion.
We next examine the electromagnetic energy supply accompanying this expansion, illustrated in panel (c) of Figure~\ref{fig:eruption-overview}.
This transport connects the declining off-equatorial horizon field to the equatorial activity.

\begin{figure}
  \centering
  \includegraphics[width=1.0\linewidth]{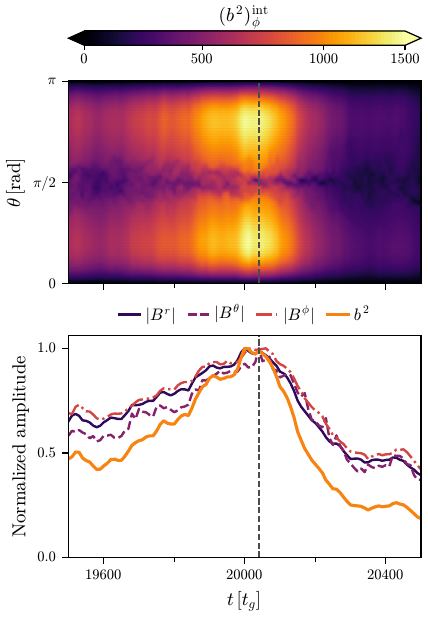}
  \caption{Top: temporal and spatial evolution of the $\phi$-integrated magnetic field strength$(b^2)^{\mathrm{int}}_\phi$, analogous to Figure~\ref{fig:horizon-evolution}.
    The dashed line marks $t = 20040\,t_g$.
    Bottom: time evolution of the horizon-field components (absolute values, normalized to their maxima).
    All components track $b^2$ closely.}
  \label{fig:horizon-b2-bcomponent}
\end{figure}

We first examine how the magnetic-field strength changes as the off-equatorial horizon flux declines.
To do so, we examine the $\phi$-integrated field strength $(b^2)^{\mathrm{int}}_\phi$ and the horizon-field components.
Figure~\ref{fig:horizon-b2-bcomponent} presents the $\phi$-integrated magnetic field strength $(b^2)^{\mathrm{int}}_\phi$ (top panel) and the horizon-field components (bottom panel).
The dashed line marks $t = 20040\,t_g$.
The evolution of $(b^2)^{\mathrm{int}}_\phi$ closely tracks that of the magnetic flux, displaying a pronounced decline after $t = 20040\,t_g$.
All field components computed from absolute values to remove sign reversals exhibit a similar temporal pattern.
The corresponding Pearson correlation coefficients with $b^2$ are 0.994 for $|B^r|$, 0.974 for $|B^\theta|$, and 0.988 for $|B^\phi|$.
The horizon-threading flux and the total field strength thus decline together, indicating a coherent weakening of the field during the eruption.
The weakening is again concentrated in the off-equatorial regions, matching the latitude pattern in Figure~\ref{fig:horizon-evolution}.
The change in magnetic energy motivates us to trace the flux decline using the electromagnetic energy flux.

\begin{figure*}[t]
  \centering
  \includegraphics[width=\textwidth]{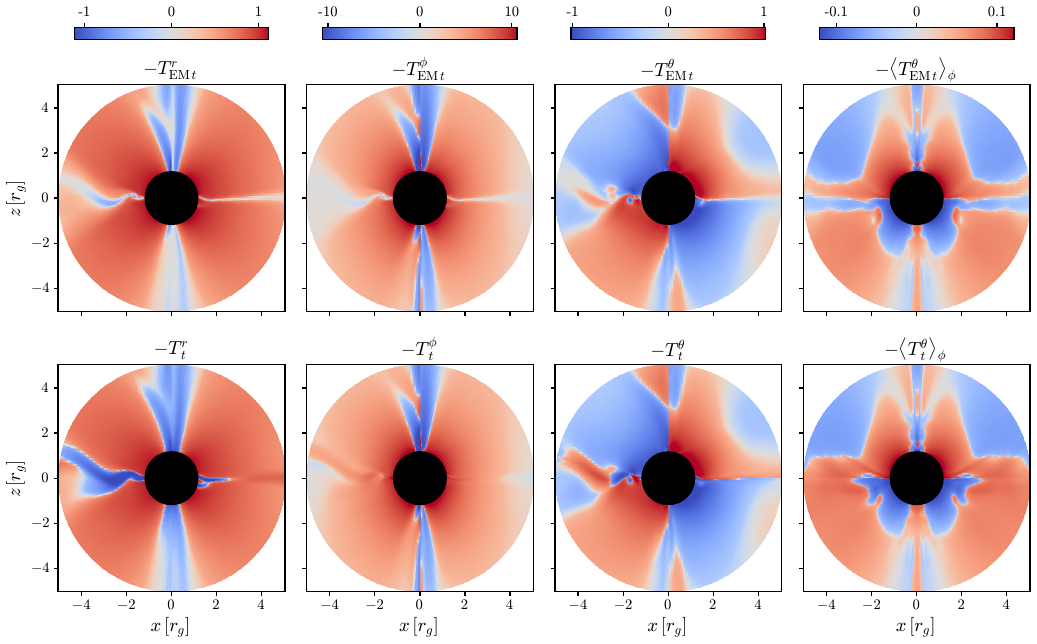}
  \caption{Components of the energy flux $-T^i_{\,t}$ at $t = 20160\,t_g$, the peak of the dimensionless magnetic flux.
  Top row: electromagnetic energy flux; bottom row: total energy flux.
  Both rows in each column share the same symmetric logarithmic color scale.
  From left to right: radial flux $-T^r_{\,t}$, azimuthal flux $-T^{\phi}_{\,t}$, polar flux $-T^{\theta}_{\,t}$, and $\phi$-averaged polar flux.
  In the first three columns, the right half of each panel is the slice at $\phi = \pi/4$ intersecting the magnetic bubble, where the polar energy flux is directed toward the equatorial plane; the left half corresponds to a typical accretion flow region without a bubble, where the $\theta$-directed flux moves away from the equator.
  The $\phi$-averaged polar flux (last column) is also predominantly equatorward, showing that the bubble-associated transport redistributes electromagnetic energy toward the equatorial region.}
  \label{fig:energy-flux-t20160}
\end{figure*}

Figure~\ref{fig:energy-flux-t20160} shows the energy-flux components at $t = 20160\,t_g$, near the middle of the eruption.
The top row gives the electromagnetic component and the bottom row the total flux.
The first three columns show $-T^r_{\,t}$, $-T^\phi_{\,t}$, and $-T^\theta_{\,t}$; the final column shows the $\phi$-averaged polar flux.
In the sliced panels, the right half cuts through the bubble at $\phi=\pi/4$, whereas the left half shows the azimuthally opposite sector, where the flow is closer to the usual accretion pattern.
For orientation, the radial and azimuthal components mostly show the usual accretion-flow pattern.
The radial flux is outward in the wind and inward in the disk, while the azimuthal flux follows disk rotation.
These components mainly identify the background flow.

The key bubble-specific signature appears in the polar component.
In the bubble sector (right half of the third column), the polar energy flux near the equator in both hemispheres points toward the equatorial plane.
It carries energy from the off-equatorial regions toward the equatorial disk.
On the opposite side, where no comparable bubble is present, the $\theta$-directed flux points away from the equator.
This asymmetry associates the expanding bubble with a distinct energy-transport pattern: electromagnetic energy flows from the off-equatorial regions toward the equator in the bubble sector, while transport in the opposite sector is poleward.
The $\phi$-averaged polar flux in the fourth column also remains predominantly equatorward, indicating that the bubble-associated redistribution makes an important contribution to the polar energy transport during the flux eruption.

\begin{figure*}[t]
  \centering
  \includegraphics[width=\textwidth]{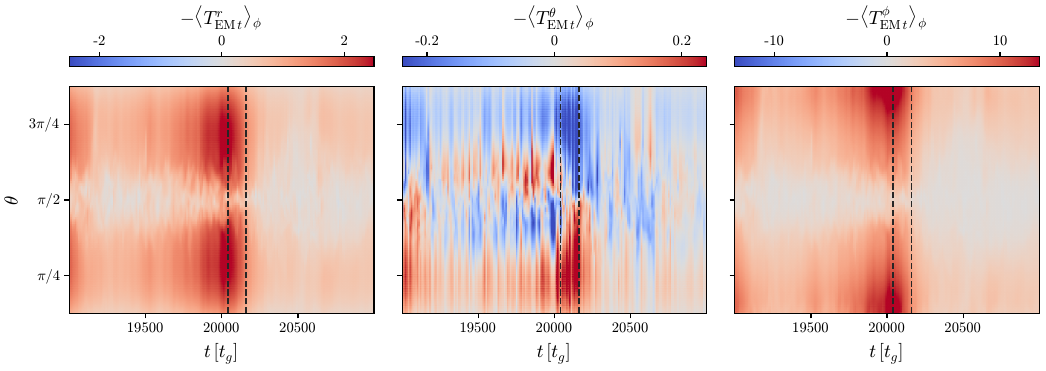}
  \caption{Time evolution of the $\phi$-averaged electromagnetic energy flux at $r=1.5\,r_g$ as a function of polar angle $\theta$ and time.
  From left to right, we show the radial electromagnetic flux $-\langle T^r_{\mathrm{EM}\,t}\rangle_\phi$, the polar electromagnetic flux $-\langle T^{\theta}_{\mathrm{EM}\,t}\rangle_\phi$, and the azimuthal electromagnetic flux $-\langle T^{\phi}_{\mathrm{EM}\,t}\rangle_\phi$.
  The two vertical dashed lines indicate the onset of the flux eruption at $t=20040\,t_g$ and the peak of the dimensionless magnetic flux at $t=20160\,t_g$.
  The polar electromagnetic flux $-\langle T^\theta_{\mathrm{EM}\,t}\rangle_\phi$ shows a clear transition from being directed away from the equatorial plane before the eruption to being directed toward the equatorial plane during the eruption, consistent with the behavior observed in Figure~\ref{fig:energy-flux-t20160}.}
  \label{fig:energy-flux-theta-time}
\end{figure*}

The time evolution of the $\phi$-averaged electromagnetic energy-flux components as a function of polar angle at $r=1.5\,r_g$ is shown in Figure~\ref{fig:energy-flux-theta-time}.
The radial electromagnetic flux is outward in the wind region, strengthens from about $t \sim 19700\,t_g$, and then weakens after $t = 20040\,t_g$, following the decline of the horizon field.
The azimuthal component is positive, indicating co-rotation with the disk, and it also strengthens during the buildup phase and weakens after the eruption.

The polar electromagnetic energy flux exhibits a clear temporal reversal.
Before the eruption, electromagnetic energy is transported from the equatorial region toward off-equatorial regions.
Around $t = 20040\,t_g$, this direction reverses, with electromagnetic energy flowing from off-equatorial regions toward the equatorial plane.
Figure~\ref{fig:energy-flux-t20160} provides the spatial context for interpreting this reversal: the equatorward polar flux is concentrated in the bubble sector, whereas the flux in the opposing, ordinary accretion-flow sector remains directed away from the equator.
The spatial contrast and temporal reversal together link bubble development to the equatorward electromagnetic energy supply.

One possible explanation for this reversal is a reduction in total pressure near the equator.
Reconnection can reduce magnetic support, while the associated outflow and bubble expansion carry plasma and energy outward.
Together, these processes may allow the surrounding magnetized plasma to move toward the equatorial reconnection region, supplying electromagnetic energy from the off-equatorial regions.

This energy transfer can contribute to the weakening of the off-equatorial magnetic field while sustaining localized reconnection near the equatorial plane.
 In this picture, it provides a possible physical link between off-equatorial field weakening and equatorial reconnection.
 Reconnection can convert part of the supplied magnetic energy into plasma internal energy, heating the plasma.
This sustained energy supply is consistent with the growth of the heated region from localized patches near the eruption onset to a more extended region as the bubble develops (Figure~\ref{fig:magnetic_bubble}).

In addition to heating the plasma, reconnection may promote bubble expansion by producing strongly curved magnetic field lines near the reconnection site.
The resulting curvature can enhance magnetic tension, helping to pull the reconnected field lines outward and thereby supporting further bubble expansion.
This mechanism may account for part of the localized outward magnetic tension identified in Section~\ref{subsec:dynamic-of-magnetic-bubble}.
The associated outflow and expansion may help maintain the pressure reduction near the equator, allowing continued energy supply from the surrounding magnetosphere as the eruption develops.

\section{Conclusions and Discussion} \label{sec:discussion}

Using high-resolution GRMHD simulations of a magnetically arrested disk, we investigate the mechanism of magnetic eruption.
The horizon-threading magnetic flux declines mainly in off-equatorial regions, whereas magnetic reconnection is concentrated near the equatorial plane.
Weak outward motion and localized hot patches are already present near the onset of the flux decline.
The heated region subsequently becomes more extended as the eruption develops.

Our interpretation connects magnetic-flux accumulation, localized reconnection, and magnetic bubble expansion.
Before the eruption, magnetic flux accumulates at the horizon, and the accretion flow already exhibits a filamentary structure with low-density channels. This morphology is consistent with an interchange interpretation. Magnetic compression from the
off-equatorial regions may facilitate reconnection.
Reconnection  can produce curved magnetic field lines whose tension can contribute to outward motion near the equator, while magnetic-pressure gradients provide more broadly distributed support for bubble expansion.
As the bubble expands, electromagnetic energy flows from the off-equatorial magnetosphere toward the equatorial region, potentially supplying energy for continued reconnection and plasma heating.
This energy transfer can contribute to the weakening of the off-equatorial magnetic field, linking the decline of horizon-threading magnetic flux to localized reconnection near the equator.

This picture complements reconnection-based interpretations by examining the dynamics and energy supply of the developing eruption.
The early  heating signatures are consistent with reconnection-based scenarios~\citep{Ripperda2022ApJ...924L..32R,Chatterjee2022ApJ...941...30C,Nalewajko2024A&A...692A..37N,Dimitropoulos2025A&A...696A..36D}; our analysis  characterizes the magnetic forces and equatorward energy transport  accompanying bubble growth and  equatorial activity.
It also complements flux-transport models for the cyclic accumulation and release of horizon-threading flux~\citep{Jacquemin-Ide2025arXiv251025842J} by resolving the bubble dynamics and the associated energy-transport pathway.

The proposed coupling between reconnection, bubble expansion, and energy transport may operate during recurrent MAD flux eruptions reported in previous studies~\citep{Dexter2020MNRAS.497.4999D,Antonopoulou2025A&A...696A..10A,Jacquemin-Ide2025arXiv251025842J}.
Following the release of accumulated magnetic flux, renewed inward flux transport may replenish the horizon-threading field and allow another eruption cycle.
Inward magnetic-flux transport is one possible influence on recurrence~\citep{Dexter2020MNRAS.497.4999D,Jacquemin-Ide2025arXiv251025842J}, but its role relative to local plasma conditions and variability in the surrounding accretion flow remains unresolved.
Our analysis does not identify the trigger of reconnection or the process that sets the recurrence timescale.
The implications of this coupling for quasi-periodic variability and recurrent flaring in systems such as Sgr~A* require further investigation~\citep{Porth2021MNRAS.502.2023P,Cemeljic2022ApJ...933...55C,Ripperda2022ApJ...924L..32R,Scepi2022MNRAS.511.3536S,Antonopoulou2025A&A...696A..10A}.

The expanding magnetic bubble associated with a magnetic eruption may be closely related to the formation and ejection of flux ropes, a phenomenon first studied in analytic models and later demonstrated in  GRMHD simulations~\citep{Yuan2009MNRAS.395.2183Y,Cemeljic2022ApJ...933...55C}.
These two studies focus on the reconnection driven by differential rotation and turbulent motion of the accretion flow.
In the picture considered here, reconnection may promote magnetic bubble expansion and continue as the eruption develops.
Bubble expansion may perturb surrounding magnetic structures and facilitate additional reconnection, potentially contributing to flux-rope formation and ejection.
Recurrence timescales provide a further point of comparison.
In the MAD simulation analysed by \citet{Cemeljic2022ApJ...933...55C}, flux-rope formation and ejection recur quasi-periodically on timescales of approximately  $1000\,t_g$.
This period is comparable to the reported period of magnetic eruption~\citep{Antonopoulou2025A&A...696A..10A,Jacquemin-Ide2025arXiv251025842J}.
\citet{Cemeljic2022ApJ...933...55C} interpret these flux-rope ejections in terms of reconnection driven by differential rotation and turbulent motions, rather than explicitly identifying them with MAD flux eruptions.
The comparable recurrence timescales suggest a possible physical connection between these phenomena.
In the context of flux-rope ejection, \citet{Lin2024MNRAS.531.3136L} modeled the reconnection-driven acceleration and subsequent radiation of electrons and showed that this process can naturally account for various observational properties of Sgr A* flares detected by the GRAVITY instrument, including the flare light curve and polarization signatures, particularly the trajectory of the ``hot spots'' detected during the flare and their super-Keplerian motion.

\begin{acknowledgments}
  The authors thank the anonymous referee for constructive comments that improved this manuscript, and colleagues and collaborators for helpful discussions.
  G.Q.\ Zhang is supported by the CPSF under Grant Numbers 2025M783224 and GZC20252102.
  F.\ Yuan is supported in part by NSFC under Grant Numbers 12133008, 12192220, 12192223, and 12361161601.
\end{acknowledgments}

\bibliography{main}
\bibliographystyle{aasjournalv7}

\end{document}